\documentclass[12pt]{article}

\usepackage{epsfig,latexsym,amsfonts,amsmath,amsthm,amssymb, mathrsfs,amsbsy,multirow,slashed,wasysym,textcomp,subfigure,hyperref,wrapfig,datetime,comment,mathtools,cancel,cite}
\usepackage[font={footnotesize,sl},bf]{caption}
\usepackage{comment}

\def\be{\begin{equation}}
\def\ee{\end{equation}}
\def\bea{\begin{eqnarray}}
\def\eea{\end{eqnarray}}
 \newcommand{\badat}{\begin{alignedat}}
 \newcommand{\eadat}{\end{alignedat}}

\long\def\new#1\endnew{{\bf #1}}		
\long\def\del#1\enddel{}

\def\del{\partial}

\usepackage{xcolor}
\definecolor{oldmauve}{rgb}{0.4, 0.19, 0.28}
\definecolor{pansypurple}{rgb}{0.47, 0.09, 0.29}
\definecolor{burgundy}{rgb}{0.5, 0.0, 0.13}
\definecolor{carminepink}{rgb}{0.92, 0.3, 0.26}
\definecolor{blue(pigment)}{rgb}{0.2, 0.2, 0.6}
\definecolor{darkseagreen}{rgb}{0.56, 0.74, 0.56}
\definecolor{darkspringgreen}{rgb}{0.09, 0.45, 0.27}
\definecolor{ceruleanblue}{rgb}{0.16, 0.32, 0.75}

\begin{document}
%\title{}
%\date{}
\numberwithin{equation}{section} % equation numbers follow sections

\begin{titlepage}
  \thispagestyle{empty}
  
  \begin{center} 
  \vspace*{3cm}
{\LARGE{Liouville strings in AdS$_3$: the worldsheet story}}

\vskip1cm

\centerline{Gaston Giribet$^1$, Pedro Schmied$^2$}
%\footnote{gg1043@nyu.edu}
%\footnote{pedroschmied0811@gmail.com}
   
\vskip1cm

$^1${Center for Cosmology and Particle Physics\\ Department of Physics, New York University\\ {\it 726 Broadway, New York City, NY10003, USA.}}\\
\vskip0.5cm

$^2${Departamento de F\'{\i}sica, Universidad de Buenos Aires, FCEN-UBA\\
Instituto de F\'{\i}sica de Buenos Aires, IFIBA-CONICET\\
{\it Ciudad Universitaria, pabell\'on 1, 1428, Buenos Aires, Argentina.}}\\

\end{center}

\vskip1cm

\begin{abstract}
The worldsheet $\sigma$-model of strings on AdS$_3$ with NS-NS fluxes can be equivalently described in terms of a Liouville field theory coupled to a timelike field with background charge, plus a marginal deformation. The operator that produces the deformation is non-normalizable and, from the string theory perspective, is associated to the spectral flow sector $\omega =2$. Its role is to control the winding number violation in scattering amplitudes. In the semiclassical (large $k$) limit, the central charge of the Liouville factor tends to that of the dual CFT$_2$, namely $c\simeq 6k$. This may appear very similar to what occurs in the Liouville-type description of the dual deformed orbifold CFT$_2$, where a twist-2 deformation operator dressed with a non-normalized field also appears. Indeed, there are some similarities to that; however, there are also significant differences, which we discuss.
\end{abstract}

\end{titlepage}

\section{Introduction}

For over twenty-five years, string theory in AdS$_3$ has served as a laboratory for investigating string theory in curved, non-compact spacetimes within the finite $\alpha'$ regime. In particular, this has enabled the investigation of the AdS/CFT holographic correspondence in a specific setting, extending beyond the supergravity limit \cite{Giveon:1998ns, deBoer:1998gyt, Kutasov:1999xu}.

One of the advantages of string theory in AdS$_3$ with NS-NS backgrounds is that it provides direct access to the worldsheet CFT formulation. This is because, as is well known, the worldsheet theory is described by the Wess-Zumino-Witten (WZW) model on $SL(2,\mathbb{R})$. This has made it possible to study the string spectrum \cite{MO, Maldacena:2000kv} and string interactions \cite{MO3} exactly.

Among the most interesting applications of string theory in AdS$_3$ is the explicit check of the AdS/CFT correspondence beyond the field theory approximation. By focusing on the calculation of protected observables, such as the 3-point correlation functions of $\frac 12$ BPS states, an exact match between observables has been verified in a concrete example where both sides of the duality are known and exactly solvable \cite{Gaberdiel:2007vu, Dabholkar:2007ey, Giribet:2007wp, Cardona:2009hk}.

Shortly thereafter, attention shifted to a specific point in the moduli space of the theory, namely the so-called tensionless point \cite{Gaberdiel:2017oqg}. This special point, corresponding to the limit in which the string length $\sqrt{\alpha '}$ equals the radius of AdS$_3$, had already been identified in the literature as a point where the theory exhibits peculiar properties \cite{Giveon:2005mi}. The formulation of the theory at this special point enabled precision tests of AdS/CFT, including the determination of the full spectrum in a specific setup \cite{Gaberdiel:2018rqv, Giribet:2018ada}. In this construction of a dual theory in terms of the symmetric orbifold, a non-compact CFT with central charge $c=1$ plays an essential role. This gave rise to further interesting generalizations. Recent works in this research line are \cite{Eberhardt, Eberhardt:2019ywk, Eberhardt:2020akk, Gaberdiel:2020ycd, Gaberdiel:2023lco, Dei:2020zui, Dei:2021xgh, Dei:2021yom, Dei:2022pkr, Balthazar:2021xeh}.

In recent attempts to construct a dual theory beyond the tensionless point, a key role is played by Liouville field theory \cite{Eberhardt}. In \cite{Eberhardt}, the authors showed that the CFT$_2$ dual to superstrings on AdS$_3\times S^3\times T^4$ for generic $k$ units of NS-NS flux is the symmetric orbifold of $\mathcal{N}=4$ super-Liouville theory times the theory on $T^4$. For the case $k=1$, which corresponds to the tensionless point, the Liouville factor of the theory disappears, leaving the symmetric orbifold of $\mathcal{M}=T^4$ and a $c=1$ direction. This gave further support to previous results in the literature. A similar Liouville construction applies to the bosonic string on AdS$_3\times \mathcal{M}$ \cite{Eberhardt}. Liouville theory also appears in discussions related to the orbifold theory in references \cite{Dei:2021xgh, Dei:2021yom, Dei:2022pkr, Balthazar:2021xeh, Dei:2020zui, Eberhardt, Eberhardt2, 231205317, Kutasov}. Of particular interest is reference \cite{Eberhardt2}, in which the author constructs a CFT dual to string theory on AdS$_3\times \mathcal{M}$, based on a symmetric orbifold of a linear dilaton theory deformed by a marginal operator from the twist-2 sector. It was already observed in \cite{Balthazar:2021xeh} that superstring theory on AdS$_3\times \mathcal{M}$ with $k<1$ turns out to be dual to a symmetric product CFT deformed by an operator in the twist-2 sector. This was generalized in \cite{Kutasov} to $k\geq 1$, with results in agreement with the analysis performed in \cite{Eberhardt2}. The authors of \cite{Kutasov} also showed that the deformation of the orbifold theory remains non-trivial at the tensionless point $k=1$.

Here, we will also present a relation between the string theory on AdS$_3\times \mathcal{M}$ and Liouville field theory; however, it is a relation of a different nature from the one considered in the context of the orbifold CFT$_2$. Our analysis will be limited entirely to the worldsheet CFT$_2$. We will propose an alternative worldsheet description of string theory on AdS$_3$ with NS-NS fluxes in terms of $c\geq 25$ Liouville theory coupled to a timelike field and deformed by a marginal operator consisting of a non-integrable Liouville operator. The comparison with the Liouville theory of the orbifold construction will be emphasized. 

The paper is organized as follows: In section 2, we will introduce the worldsheet theory and discuss its properties. We present the notation, the form of the action, the spectrum and representations, and the explicit form of the vertex operators. In section 3, we will compute the correlation functions of the proposed worldsheet theory, and in section 4 we will show that the integrals of those correlation functions exactly reproduce the string scattering amplitudes in AdS$_3$, with the winding string sectors included. In section 5, we will discuss the role of the marginal operators appearing in the proposed worldsheet theory; in particular, we will explain that one such operator controls the violation of the winding number conservation in AdS$_3$. We will also compare the marginal operators of the worldsheet theory with those proposed in the dual description. We will conclude in section 6 with some final remarks and open questions.

\section{Conformal field theory}
\subsection{The worldsheet theory}

In this section, we will construct a worldsheet theory based on Liouville theory coupled to other fields. Let us begin by considering the following Lagrangian for the free theory
\begin{equation}\label{L0}
\mathcal{L}_0= (\partial \phi_+)^2+ \sqrt{2} \,Q_+\,R\,\phi_+- (\partial \phi_-)^2+ \sqrt{2}\, Q_-\,R\,\phi_-+ (\partial X)^2
\end{equation}
with background charges
\begin{equation}
Q_+=b+\frac 1b=\sqrt{k-2}+\frac{1}{\sqrt{k-2}}, \ \ \ \ \ \ Q_-=\sqrt{k}\, ,
\end{equation}
where, for convenience, we define $b=1/\sqrt{k-2}$. Our notation is such that $(\partial \phi_{\pm})^2=\partial \phi_{\pm}\bar \partial \phi_{\pm}$, with $\partial=\partial_{z}$, $\bar \partial=\partial_{\bar z}$, $g_{z\bar z}=g_{\bar z z}=\frac 12 $.

The field content corresponds to a $c_+\geq 25$ spacelike field $\phi_+$ with background charge $Q_+$, a $c_-<1$ timelike field $\phi_-$ with background charge $Q_-$, and a free spacelike scalar boson $X$ with $c_X=1$. The background charges satisfy the relation
\begin{equation}
Q_+^2=b^2+Q_-^2=\frac{\,\,(k-1)^2}{{k-2}}.\label{antote}
\end{equation}
This yields the total central charge
\begin{equation}
c=c_++c_-+c_X=1+6Q_+^2+1-6Q^2_-+1=
13+6b^{2}+\frac{6}{b^2}+1-6k+1
=3+\frac{6}{k-2}
\end{equation}
which coincides with the central charge of the level-$k$ WZW model on $SL(2,\mathbb{R})$; namely
\begin{equation}
c_{SL(2,\mathbb{R})}=\frac{3k}{k-2}
\end{equation}
It is worth noting that the Liouville central charge $c_+ = 1+6k+6/(k-2)$ tends to $c_+\simeq 6k$ in the semiclassical limit $k\to \infty $. In that limit, $c_-=1-6k$ becomes infinitely negative.

We will now add to the Lagrangian (\ref{L0}) the following interaction term
\begin{equation}
\mathcal{L}_{\text{I}} = 2\pi \,e^{\sqrt{2}b\phi_+}-{2\pi \lambda}\, e^{-\frac{\phi_+}{\sqrt{2}b}-\sqrt{\frac{k}{2}}\phi_-}\label{L26}
\end{equation}
with $\lambda$ being a constant. Recall: $k=2+b^{-2}$. The operator (\ref{L26}) introduces a self-interaction for $\phi_+$, while coupling $\phi_-$ to $\phi_+$.

The theory defined by Lagrangian density $\mathcal{L}_0+\mathcal{L}_{\text{I}}$ turns out to be an interacting 2-dimensional CFT of the form
\begin{equation}
\text{Liouville}\,\,\times \,\, i\mathbb{R}_{Q_-} \,\otimes\,\, U(1)\label{dUno}
\end{equation}
deformed by the marginal operator 
\begin{equation}
\mathcal{O}(z,\bar z)=e^{-\frac{\phi_+(z,\bar z)}{\sqrt{2}b}-\sqrt{\frac{k}{2}}\phi_-(z,\bar z)}\label{dFinal}.
\end{equation}
This operator is central to our analysis; we will analyze its properties in detail in section \ref{Section5}.

Now, let us discuss the vertex operators. The first factor in (\ref{dUno}) is a $c_+\geq 25$ Liouville theory, the second factor is a timelike free boson with a (real) background charge $\sqrt{k}$, and the third is a free boson. In such a CFT$_2$, we consider vertex operators of the form
\begin{equation}
V_{j,m,\bar{m}}^{\omega }(z,\bar z) = N_{j,m,\bar{m}}\, e^{\sqrt{2}\alpha _+\phi_+(z,\bar z)+\sqrt{2}\alpha_-\phi_-(z,\bar z)+\sqrt{2}p\,X(z,\bar z)}\label{vertex}
\end{equation}
with momenta $ip_{\phi_+}=\alpha_+$, $ip_{\phi_-}=\alpha_-$, $ip_X=p$, which can be written as
\begin{equation}
\alpha_+=b\Big(j+\frac k2\Big),\ \ \ \ \alpha_-=\frac{\sqrt{k}}{2}-\frac{m}{\sqrt{k}}, \ \ \ \ p=\frac{m}{\sqrt{k}}+\frac{\sqrt{k}\omega}{2} 
\end{equation}
and a normalization factor
\begin{equation}
N_{j,m,\bar m} = \frac{ \, \Gamma(-j-m)}{\pi^2\lambda\,\Gamma(1+j+\bar m)}.
\end{equation}
This normalization factor satisfies $N_{j,m,\bar m}N_{-1-j,-\bar m,-m}=\pi^{-4}\lambda^{-2}$, $N_{j,m,\bar m}=(-1)^{2j+1+m+\bar m}N_{j,\bar m, m}$, $N_{j,m,\bar m}=(-j-m)N_{j,m+1,\bar m}$.

Operators (\ref{vertex}) create primary states of momenta $\alpha_+$, $\alpha_-$, $p$. Correspondingly, we have $\bar{\alpha}_-={\sqrt{k}}/{2}-{\bar{m}}/{\sqrt{k}}$, since $\alpha_-\phi_-(z,\bar{z})$ has actually to be understood as $\alpha_-\phi_-(z)+\bar\alpha_-\bar{\phi}_-(\bar{z})$, as usual with a chiral momentum. Analogously, we have $\bar p=(\bar m+\frac k2 \omega)/\sqrt{k}$. So, we have
\begin{equation}
\frac{\alpha_-+p}{\sqrt{2}} = \frac{\bar{\alpha}_-+\bar{p}}{\sqrt{2}} = \sqrt{\frac k2}\, (\omega + 1).
\end{equation}
In terms of these variables, the conformal weights of the operators (\ref{vertex}) read
\begin{equation}
\Delta_{j,m,\omega} = \alpha_+(Q_+-\alpha_+)-\alpha_-(Q_--\alpha_-)-p^2=-\frac{j(j+1)}{k-2}-m\omega -\frac{k}{4}\omega^2,
\end{equation}
together with its conjugate counterpart $\bar\Delta_{j,\bar m,\omega}$. This can easily be verified by computing the operator product expansion (OPE) between the stress tensor derived from (\ref{L0}) and the vertex (\ref{vertex}). Then, the Virasoro constraint $L_0\approx 1$ demands the level-$N$ excited string states on AdS$_3\times \mathcal{M}$ to obey
\begin{equation}
m+ \frac k2\, \omega \, =\, \frac k4 \,\omega +\frac{1}{\omega }\, \Big(N-1-\frac{j(j+1)}{k-2}+h_{\mathcal{M}}\Big)
\end{equation}
from which the energy spectrum may be obtained; see below. $h _{\mathcal{M}}$ is the conformal dimension on the CFT$_2$ on the internal space $\mathcal{M}$.

Momenta $\alpha_{\pm}$ and $p$ are written in terms of labels $j,m,\bar m$ and $\omega$, which organize in the representations of the universal covering of $SL(2,\mathbb{R})$ that are relevant to describe the string states in AdS$_3$ \cite{MO}. $m-\bar{m}$ is the spin of the state in AdS$_3$, $m+\bar{m}+k\omega$ is the energy, and $\omega $ is the winding number. The isospin label $j$ gives the radial momentum. For example, normalizable operators in the spacelike Liouville theory correspond to $\alpha_+\in \frac{Q_+}{2}+i\mathbb{R}$, and these values correspond to labels $j\in -\frac 12 +i\mathbb{R}$ of the continuous principal series of $SL(2,\mathbb{R})$, describing long strings in AdS$_3$ (also note that the Weyl reflection $j\to -1-j$, which leaves $\Delta_{j,m,\omega }$ invariant, corresponds to the Liouville reflection $\alpha_+\to Q_+-\alpha_+$). Short strings, in contrast, have $j\in \mathbb{R}$, represent bound states, and correspond to non-normalizable operators ($\alpha_+\in \mathbb{R}$) of Liouville theory. 

No-ghost theorem in AdS$_3$ demands the $SL(2,\mathbb{R})_k$ discrete representations describing short string states to obey the following bound on the radial momentum \cite{MO}
\begin{equation}
\frac{1-k}{2}<j<-\frac 12\, .
\end{equation}
While the upper bound (lower bound in the conventions of \cite{MO}) translates into the well-known Seiberg bound on the Liouville momentum of the corresponding field, namely 
\begin{equation}
\alpha_+<\frac{Q_+}{2}\,, 
\end{equation}
the lower bound on $j$ corresponds to the condition $\alpha_+>b/{2}$, which would be interesting to understand from the Liouville theory perspective. For the timelike field $\phi_-$, a similar bound exists: requiring $\alpha_-<{Q_-}/{2}$ translates into the positivity condition $m>0$. Changing $m,\omega \to -m,-\omega$ corresponds to performing $\alpha_- ,p \to Q_--\alpha_-, -p$. In fact, the positivity of energy in AdS$_3$, namely $m+\frac k2 \omega\geq 0 $, translates into $p\geq 0$. Here, without loss of generality, we will focus on the cases in which momenta obey $m >0$, $\bar m >0$ and $\omega <0$, but analogous formulae hold for the $\mathbb{Z}_2$-reflected (time-reversed) case. 

Now, let us discuss the symmetry and currents. In terms of fields $\phi_+$, $\phi_-$ and $X$ the local currents $J^{\pm , 3}(z)$ that generate the affine Kac-Moody $\hat{sl}(2)_k$ symmetry take a form similar to that studied in \cite{G}. The Cartan current $J^3(z)$, for example, takes the simple form
\begin{equation}
J^3(z) \,=\, -\sqrt{\frac{k}{2}}\partial X(z),
\end{equation}
which yields the OPE
\begin{equation}
J^3(z)\,J^3(0) \, = \,\frac k2 \,  \partial X(z)\, \partial X(0)\, =\, - \frac{k}{2\,z^2} \, + \, ...   
\end{equation}
where the ellipsis stands for regular terms. The action on the vertex operators is
\begin{equation}
J^3(z)\,V_{j,m,\bar m}(0)\, =\, \frac{m+\frac k2 \omega }{z}\, V_{j,m,\bar m}(0)\, +\, ...\label{J3}
\end{equation}

In order to have a clearer spacetime interpretation of the fields, it is convenient to perform the double Wick rotation, namely
\begin{equation}
X(z,\bar z)\,\to \,i\,t(z,\bar z)\, , \ \ \ \ \ \phi_-(z,\bar z)\,\to \,i\,\theta(z,\bar z)\, , \ \ \ \ \ \phi_+(z,\bar z)\,\to \,\phi(z,\bar z)\, .\label{J}
\end{equation}
In this case, the free Lagrangian reads
\begin{equation}
\mathcal{L}_0= - (\partial t)^2
+ (\partial \theta)^2
+(\partial \phi)^2+ \sqrt{2}\, Q_{\phi}\,R\,\phi + i\sqrt{2k} \,R\,\theta \label{Lposta}
\end{equation}
with a background charge
\begin{equation}
Q_{\phi} \, = \, \frac{1}{\sqrt{k-2}}+\sqrt{k-2}
\end{equation}
plus an imaginary contribution to the worldsheet action $\sqrt k \int d^2zR\theta $. The imaginary background term is reminiscent of the one appearing in the free field realization of the Minimal Models \cite{Fateev, Fateev2}. The variable $\theta $ can be thought of as an angular direction; in fact, for $\alpha '=1$ the quantity $2\pi \sqrt k \theta $ is the boundary perimeter. This might be useful thinking of a classical long string solution at fixed radius $\langle \phi_+ \rangle $ and extending along the worldsheet coordinate $\sigma ^0=\tau = t$ and $\sigma ^1=\sigma =  \sqrt k \theta $.

In the Wick rotated coordinates (\ref{J}) the vertex operators take a more familiar form 
\begin{equation}
V_{j,m,\bar m }(z,\bar z ) \propto e^{i\sqrt{2}p_{\theta }\theta (z,\bar z )  }e^{i\sqrt{2}p_{0}t (z,\bar z )  },
\end{equation}
with $p_{\theta}=\alpha_-$, $p_0=p$. The interpretation of $t$ as a time and $\theta $ as an angle is compatible with the signs of the kinetic terms of these fields in (\ref{Lposta}). Now, the $U(1)$ Cartan current is $J^3(z)= -i \sqrt{\frac k2}\partial t(z)$ and represents the energy $p_0=m+\frac k2 \omega $. According to this, $\phi(z)$ corresponds to the radial direction, $\theta $ to the angular direction of the AdS$_3$ boundary and $t $ to time.

\subsection{Liouville field theory}

In terms of the fields $\phi_+,\phi_-, X$, the full CFT worldsheet action is
\begin{equation}
S[\phi_+,\phi_-,X] = \frac{1}{2\pi }\int d^2z \, (\mathcal{L}_{0} + \mathcal{L}_{\text{I}}) \label{Uno}
\end{equation}
We will find it convenient to write it as the sum of the three terms
\begin{eqnarray}
S[\phi_+] &=& + \frac{1}{2\pi}\int d^2z \, \Big(  (\partial \phi_+)^2+\sqrt{2}\,Q_+R\phi_+ +2\pi \, e^{\sqrt{2}b\phi_+} \Big)\\
S[\phi_-] &=& -\frac{1}{2\pi}\int d^2z \, \Big(  (\partial \phi_-)^2-\sqrt{2}\,Q_-R\phi_- \Big)\\
S[X] &=& + \frac{1}{2\pi}\int d^2z \,  (\partial X)^2
\end{eqnarray}
plus the deformation operator
\begin{equation}
S_{\mathcal{O}}[\phi_+,\phi_-]= -\lambda \int d^2z \, e^{-\frac{\phi_+}{\sqrt{2}b}-\sqrt{\frac{k}{2}}\phi_-}\label{Deformation}
\end{equation}
As mentioned before, the properties of this marginal operator will be discussed in detail later.

\section{Correlation functions}

Now, let us consider the $n$-point correlation functions on the sphere; namely
\begin{eqnarray}
\left\langle \prod_{i=1}^{n}\,  V_{j_i,m_i,\bar{m}_i}^{\omega_i}(z_i,\bar{z}_i) \, \right\rangle &=&\int \mathcal{D}\phi_+\mathcal{D}\phi_-\mathcal{D}X\, e^{-S[\phi_+]-S[\phi_-]-S[X]}e^{-S_{\mathcal{O}}[\phi_+, \phi_-]} \,
\prod_{i=1}^nV_{j_i,m_i,\bar{m}_i}^{\omega_i}(z_i,\bar{z}_i)\nonumber 
\end{eqnarray}
By definition, these can be written as
\begin{eqnarray}
\left\langle \prod_{i=1}^{n}\,  V_{j_i,m_i,\bar{m}_i}^{\omega_i}(z_i,\bar{z}_i) \, \right\rangle &=&\sum_{\ell =0}^{\infty}\frac{(-1)^{\ell }}{\ell !}\int \mathcal{D}\phi_+\mathcal{D}\phi_-\mathcal{D}X\, e^{-S[\phi_+]-S[\phi_-]-S[X]}\Big( S_{\mathcal{O}}[\phi_+, \phi_-]\Big)^{\ell } \nonumber \\
&&\times \, \prod_{i=1}^n
\Big( N_{j_i,m_i,\bar{m}_i}\, e^{\sqrt{2}\alpha^{i} _+\phi_+(z_i,\bar z_i)+\sqrt{2}\alpha^i_-\phi_-(z_i,\bar z_i)+\sqrt{2}p_iX(z_i,\bar z_i)} \Big)\label{Final}
\end{eqnarray}
where, in the second line, we have used the explicit form of the vertices $V_{j_i,m_i,\bar{m}_i}^{\omega_i}$.
Using also the explicit form of $S_{\mathcal{O}}[\phi_+,\phi_-]$, this can easily be shown to take the following form
\begin{eqnarray}
&&\left\langle \prod_{i=1}^{n}\,  V_{j_i,m_i,\bar{m}_i}^{\omega_i}(z_i,\bar{z}_i) \, \right\rangle =\, \prod_{i=1}^nN_{j_i,m_i,\bar{m}_i}\,\, 
\int \mathcal{D}X\,e^{-S[X]}\, \Big(\prod_{i=1}^ne^{\sqrt{2}p_iX(z_i,\bar z_i)}\Big)\, \nonumber \\
&&  \ \ \ \ \ \ \ \ \ \ \ \ \ \ \ \ \ \ \ \ \ \ \times \,
\sum_{\ell =0}^{\infty}\frac{\lambda^{\ell }}{\ell !}\, \int \prod_{r=1}^{\ell }d^2y_r\,\Big[ \int \mathcal{D}\phi_+\, e^{-S[\phi_+]}\, \Big( \prod_{i=1}^n\, e^{\sqrt{2}\alpha^i_+\phi_+(z_i,\bar{z}_i)}\, \prod_{r=1}^{\ell}e^{-\frac{1}{\sqrt{2}b}\phi_+(y_r,\bar{y}_r)}
\Big)\nonumber \\
&&
\ \ \ \ \ \ \ \ \ \ \ \ \ \ \ \ \ \ \ \ \ \ \ \ \ \ \ \ \ \ \ \ \ \ \ \ \ \ \ \ \ \  \ \ \ \ \ \ \ \times \, \int \mathcal{D}\phi_-\, e^{-S[\phi_-]}\, \Big( \prod_{i=1}^ne^{\sqrt{2}\alpha^i_-\phi_-(z_i,\bar{z}_i)}\prod_{r=1}^{\ell}e^{-\sqrt{\frac k2}\phi_-(y_r,\bar{y}_r)}
\Big)\, \Big]\nonumber 
\end{eqnarray}

The integral over $X$ is Gaussian and can be performed explicitly. Using $p_i=(m_i+\frac k2 \omega _i)/{\sqrt{k}}$ and integrating over the zero mode $c=X-\langle X \rangle$, this yields a $\delta$-function
\begin{equation}
\delta \Big(\sum_{i=1}^n (m_i+\frac k2 \omega_i)\Big)\,\times \, \delta \Big(\sum_{i=1}^n (\bar{m}_i+\frac k2 \omega_i)\Big)
\end{equation}
which is equivalent to 
\begin{equation}
\delta_{\sum_{i}(m_i-\bar{m}_i)}\,\times \, \delta \Big(\sum_{i=1}^n (m_i+\bar{m}_i+ k \omega_i)\Big)\label{deltas}\, ,
\end{equation}
where we have used the condition $m_i-\bar{m}_i\in \mathbb{Z}$ coming from the representations of $SL(2,\mathbb{R})\times SL(2,\mathbb{R})$. Using this and recalling that $\alpha^i_-={\sqrt{k}}/{2}-{m_i}/{\sqrt{k}}$, we obtain
\begin{eqnarray}
&&\left\langle \prod_{i=1}^{n}\,  V_{j_i,m_i,\bar{m}_i}^{\omega_i}(z_i,\bar{z}_i) \, \right\rangle =\,\delta_{\sum_{i}(m_i-\bar{m}_i)}\, \delta \Big(\sum_{i=1}^n (m_i+\bar{m}_i+ k \omega_i)\Big) \,
 \nonumber \\
&& 
\ \ \ \ \ \ \ \ \  \ \ \ \ \ \ \ \ \ \ \ \ \ \times \,
\prod_{i<j}^n\Big(z_{ij}^{-\frac 2k (m_i+\frac k2 \omega_i)(m_j+\frac k2 \omega_j)} \, \bar{z}_{ij}^{-\frac 2k(\bar m_i+\frac k2 \omega_i)(\bar m_j+\frac k2 \omega_j)}\Big)\, \prod_{i=1}^nN_{j_i,m_i,\bar{m}_i}\,\nonumber \\
&&  \ \ \ \ \ \ \ \ \  \ \ \ \ \ \ \ \ \ \ \ \ \ \times \,
\sum_{\ell =0}^{\infty}\frac{\lambda^{\ell }}{\ell !}\, \int \prod_{r=1}^{\ell }d^2y_r\,\Big[ \int \mathcal{D}\phi_+\, e^{-S[\phi_+]}\, \Big( \prod_{i=1}^n\, e^{\sqrt{2}\alpha^i_+\phi_+(z_i,\bar{z}_i)}\, \prod_{r=1}^{\ell}e^{-\frac{1}{\sqrt{2}b}\phi_+(y_r,\bar{y}_r)}
\Big)\nonumber \\
&&
\ \ \ \ \  \ \ \   \ \ \ \ \ \ \ \ \ \ \ \ \ \ \ \ \ \ \ \ \ \ \ \ \ \ \ \ \ \ \ \  \ \ \ \ \ \ \ \times \, \int \mathcal{D}\phi_-\, e^{-S[\phi_-]}\, \Big( \prod_{i=1}^ne^{-\sqrt{\frac 2k}(m_i-\frac k2)\phi_-(z_i)}\prod_{r=1}^{\ell}e^{-\sqrt{\frac k2}\phi_-(y_r)}
\Big)\, 
\nonumber \\
&&
\ \ \ \ \ \ \ \ \ \ \ \ \ \ \ \ \ \ \ \ \ \ \ \ \ \ \ \ \ \ \ \ \ \ \ \ \ \ \ \ \  \ \ \ \ \ \ \ \ \ \  \ \ \ \ \ \ \ \ \ \ \ \ \ \ \  \times \, \Big( \prod_{i=1}^ne^{-\sqrt{\frac 2k}(\bar m_i-\frac k2)\bar \phi_-(\bar z_i)}\prod_{r=1}^{\ell}e^{-\sqrt{\frac k2}\bar \phi_-(\bar y_r)}
\Big)
\, \Big]\nonumber 
\end{eqnarray}
where $z_{ij}\equiv z_i-z_j$. Here, we used the spacelike free field correlator $\langle X(z_i)\,X(z_j)\rangle = -\log z_{ij}$.

Next, we can perform the integration in $\phi_-$. Such integral can also be performed explicitly as it is just a free theory with background charge. Using the timelike free field correlator $\langle \phi_-(z_i)\,\phi_-(z_j)\rangle = +\log z_{ij}$, we obtain
\begin{eqnarray}
&&\left\langle \prod_{i=1}^{n}\,  V_{j_i,m_i,\bar{m}_i}^{\omega_i}(z_i,\bar{z}_i) \, \right\rangle =\,\delta_{\sum_{i}(m_i-\bar{m}_i)}\, \delta \Big(\sum_{i=1}^n (m_i+\bar{m}_i+ k \omega_i)\Big) \,
 \nonumber \\
&& 
\ \ \ \ \ \ \ \ \  \ \ \ \ \ \ \ \ \ \ \ \ \ \times \,
\prod_{i<j}^n\Big(z_{ij}^{-\frac 2k (m_i+\frac k2 \omega_i)(m_j+\frac k2 \omega_j)} \, \bar{z}_{ij}^{-\frac 2k(\bar m_i+\frac k2 \omega_i)(\bar m_j+\frac k2 \omega_j)}\Big)\, \prod_{i=1}^nN_{j_i,m_i,\bar{m}_i}\,\nonumber \\
&&  \ \ \ \ \ \ \ \ \  \ \ \ \ \ \ \ \ \ \ \ \times \,
\sum_{\ell =0}^{\infty}\frac{\lambda^{\ell }}{\ell !}\, \int \prod_{r=1}^{\ell }d^2y_r\,\Big[ \int \mathcal{D}\phi_+\, e^{-S[\phi_+]}\, \Big( \prod_{i=1}^n\, e^{\sqrt{2}\alpha^i_+\phi_+(z_i,\bar{z}_i)}\, \prod_{r=1}^{\ell}e^{-\frac{1}{\sqrt{2}b}\phi_+(y_r,\bar{y}_r)}
\Big)\nonumber \\
&&
\ \ \ \ \ \ \ \ \  \times \, \Big( \prod_{i<j}^{n} z_{ij}^{\frac 2k (m_i-\frac k2 )(m_j-\frac k2 )} \bar{z}_{ij}^{\frac 2k (\bar m_i-\frac k2 )(\bar m_j-\frac k2 )}\Big) 
\, \Big(\prod_{i=1}^n\prod_{r=1}^{\ell } (z_i-y_r)^{m_i-\frac k2}
(\bar z_i-\bar y_r)^{\bar m_i-\frac k2}
\Big)\, \prod_{r<t}^{\ell }|y_{rt}|^{k}\Big]
\nonumber
\end{eqnarray}

By integrating over the zero mode of $\phi_-$, and taking into account the contribution of the background charge $Q_-=\sqrt k$ to it, we obtain
\begin{equation}
\sum_{i=1}^{n} \Big(m_i-\frac k2\Big)+\frac k2\, \ell +k=0, \ \ \ \ 
\sum_{i=1}^{n} \Big(\bar m_i-\frac k2\Big)+\frac k2\, \ell +k=0,
\end{equation}
that is
\begin{equation}
\sum_{i=1}^{n} m_i =\sum_{i=1}^{n} \bar m_i = \frac k2 (n-\ell-2).
\end{equation}
Then, taking into account (\ref{deltas}), it yields
\begin{equation}
\sum_{i=1}^{n}\Big(m_i-\bar{m}_i\Big)=0,\ \ \ \ 
\sum_{i=1}^{n}\omega_i=\ell -n+2 \,,\label{La35}
\end{equation}
which are, of course, compatible with
\begin{equation}
\sum_{i=1}^{n}m_i=\sum_{i=1}^{n}\bar{m}_i=-\frac k2 \sum_{i=1}^{n}\omega_i.
\end{equation}
From this, we observe that the insertion of $\ell$ operators in each step in the sum would correspond to contributions with different total number $w\equiv -\sum_{i=1}^n\omega_i$.

Reordering factors and putting all the powers of $z_{ij}$ together, the $n$-point correlator takes the form
\begin{eqnarray}
&&\left\langle \prod_{i=1}^{n}\,  V_{j_i,m_i,\bar{m}_i}^{\omega_i}(z_i,\bar{z}_i) \, \right\rangle =\,\delta_{\sum_{i}(m_i-\bar{m}_i)}\, \delta \Big(\sum_{i=1}^n (m_i+\bar{m}_i+ k \omega_i)\Big) \,
\prod_{i<j}^{n} z_{ij}^{\beta_{ij}}\bar{z}_{ij}^{\bar{\beta}_{ij}}\, \prod_{i=1}^nN_{j_i,m_i,\bar{m}_i}\,\nonumber \\
&&  \ \ \ \ \ \ \ \ \  \ \ \ \ \ \ \ \ \ \ \ \times \,
\sum_{\ell =0}^{\infty}\frac{\lambda^{\ell }}{\ell !}\, \int \prod_{r=1}^{\ell }d^2y_r\,\Big[\,\Big(\prod_{i=1}^n\prod_{r=1}^{\ell } (z_i-y_r)^{m_i-\frac k2}
(\bar z_i-\bar y_r)^{\bar m_i-\frac k2}
\Big)\, \prod_{r<t}^{\ell }|y_{rt}|^{k}\nonumber \\
&& \ \ \ \ \ \ \ \ \ \ \ \ \ \ \ \ \ \ \ \ \ \ \ \ \ \times \, 
\int \mathcal{D}\phi_+\, e^{-S[\phi_+]}\,  \prod_{i=1}^n\, e^{\sqrt{2}\alpha^i_+\phi_+(z_i,\bar{z}_i)}\, \prod_{r=1}^{\ell}e^{-\frac{1}{\sqrt{2}b}\phi_+(y_r,\bar{y}_r)}\, \Big]
\label{correlatorq}
\end{eqnarray}
with
\begin{eqnarray}
\beta_{ij}&=& \frac 2k \Big(m_i-\frac k2\Big)\Big(m_j-\frac k2\Big)-\frac 2k\Big(m_i+\frac k2 \omega_i\Big)\Big(m_j+\frac k2 \omega_j\Big) , \label{beta} \\
\bar{\beta}_{ij}&=& \frac 2k \Big(\bar m_i-\frac k2\Big)\Big(\bar m_j-\frac k2\Big)-\frac 2k\Big(\bar m_i+\frac k2 \omega_i\Big)\Big(\bar m_j+\frac k2 \omega_j\Big).\label{barbeta}
\end{eqnarray}

The third line in the expression (\ref{correlatorq}) is an $(n+\ell)$-point correlation function in the Liouville field theory (LFT),
\begin{equation}
\left\langle \prod_{i=1}^{n}\,  V^{L}_{\alpha_+^i}(z_i,\bar{z}_i) \, \prod_{r=1}^{\ell }\,  V^{L}_{-\frac{1}{2b}}(y_r,\bar{y}_r)  \right\rangle_{\text{LFT}}\label{la311}
\end{equation}
involving $\ell $ operators $V_{-\frac{1}{2b}}^{L}=e^{\sqrt{2}\alpha^+_{1,2}\phi_+(y_r,\bar y_r)}$ of special value of momentum $\alpha^+_{1,2}=-\frac{1}{2b}$. Here, $V^{L}_{\alpha}(z,\bar z)=e^{\sqrt 2 \alpha\phi_+(z,\bar z)}$. These values of momentum correspond to degenerate operators in Liouville theory that create a non-normalizable state that contains a null descendant at the second level of the Verma module. It is important to know that such a correlator may vanish for $\ell\geq n-2$. For example, this can be seen explicitly in the case of the 3-point function by evaluating the Dorn-Otto-Zamolodchikov-Zamolodchikov (DOZZ) formula \cite{DO,ZZ} in values $\alpha_1=\alpha$, $\alpha_2=\alpha_3=-\frac{1}{2b}$ and analyzing the zeros of the $\Upsilon$-function. So, this restricts the sum over $\ell$ in the general formula thereby rendering it a finite sum, now running from $\ell =0$ to $\ell = n-2$ (i.e., from $w=n-2$ to $w=0$, which is compatible with the pattern of winding number (non-)conservation in AdS$_3$, cf. Appendix D in \cite{MO3}. Using this and the explicit form of $N_{j_i,m_i,\bar m_i}$, we obtain
\begin{eqnarray}
&&\left\langle \prod_{i=1}^{n}\,  V_{j_i,m_i,\bar{m}_i}^{\omega_i}(z_i,\bar{z}_i) \, \right\rangle =\,\delta_{\sum_{i}(m_i-\bar{m}_i)}\, \delta \Big(\sum_{i=1}^n (m_i+\bar{m}_i+ k \omega_i)\Big) \,
\prod_{i<j}^{n} z_{ij}^{\beta_{ij}}\bar{z}_{ij}^{\bar{\beta}_{ij}}\, \prod_{i=1}^n\frac{\Gamma(j_i-m_i)}{\Gamma(1+j_i+\bar{m}_i)}\,\nonumber \\
&&  \ \ \ \ \ \ \ \ \  \ \ \ \ \ \ \ \ \ \ \ \times \,
\sum_{\ell =0}^{n-2}\frac{\lambda^{\ell -n}}{\ell !\pi^{2n}}\, \int \prod_{r=1}^{\ell }d^2y_r\,\Big(\prod_{i=1}^n\prod_{r=1}^{\ell } (z_i-y_r)^{m_i-\frac k2}
(\bar z_i-\bar y_r)^{\bar m_i-\frac k2}
\Big)\, \prod_{r<t}^{\ell }|y_{rt}|^{k}\nonumber \\
&& \ \ \ \ \ \ \ \ \ \ \ \ \ \ \ \ \ \ \ \ \ \ \ \ \ \times \, 
\left\langle \prod_{i=1}^{n}\,  V^{L}_{\alpha_+^i}(z_i,\bar{z}_i) \, \prod_{r=1}^{\ell }\,  V^{L}_{-\frac{1}{2b}}(y_r,\bar{y}_r)  \right\rangle_{\text{LFT}}
\label{correlatoru}
\end{eqnarray}
Relabeling as $\text{r}\equiv{n-2-\ell }$, this takes the final form 
\begin{eqnarray}
&&\left\langle \prod_{i=1}^{n}\,  V_{j_i,m_i,\bar{m}_i}^{\omega_i}(z_i,\bar{z}_i) \, \right\rangle =\,
\frac{1}{2\pi^3b\lambda^2}\,\prod_{i<j}^{n} z_{ij}^{\beta_{ij}}\bar{z}_{ij}^{\bar{\beta}_{ij}}\, \,\prod_{i=1}^n\frac{\Gamma(-j_i-m_i)}{\Gamma(1+j_i+\bar{m}_i)}\,
\sum_{\text{r} =0}^{n-2}\frac{2\pi^{3-2n}b\lambda^{-\text{r}}}{(n-2-\text{r}) !}\,  \nonumber \\
&&  \ \ \ \ \ \ \times \,
\int \prod_{a=1}^{n-2-\text{r} }d^2y_a\,\delta \Big( \sum_{i=1}^nm_i-\frac k2 \text{r}\Big)\,\delta \Big( \sum_{i=1}^n\bar{m}_i-\frac k2\text{r}\Big)\,\Big(\prod_{i=1}^n\prod_{a=1}^{n-2-\text{r} } (z_i-y_a)^{m_i-\frac k2}
(\bar z_i-\bar y_a)^{\bar m_i-\frac k2}
\Big)\nonumber \\
&& \ \ \ \ \ \ \ \ \ \ \ \ \times \,\prod_{a<a'}^{n-2-\text{r} }|y_{aa'}|^{k}\,  
\left\langle \prod_{i=1}^{n}\,\,  V^{L}_{\alpha_+^i}(z_i,\bar{z}_i) \, \prod_{a=1}^{n-2-\text{r} }\,  V^{L}_{-\frac{1}{2b}}(y_a,\bar{y}_a)  \right\rangle_{\text{LFT}}
\label{correlator}
\end{eqnarray}
with
\begin{eqnarray}
\beta_{ij}&=& \frac k2 - \frac k2 \omega_i \omega_j-\omega_im_j-\omega_jm_i -m_i-m_j\label{barbeta}\\
\bar{\beta}_{ij}&=& \frac k2 - \frac k2 \omega_i \omega_j-\omega_i\bar m_j-\omega_j\bar m_i -\bar m_i-\bar m_j \label{barbeta}
\end{eqnarray}
and
\begin{equation}
\alpha_+^i=b\Big(j_i + \frac{1}{2b^2}+1 \Big)\ , \ \ \ \ b=\frac{1}{\sqrt{k-2}}\label{alpha}
\end{equation}
Here, we have used (\ref{La35}) to rewrite the $\delta$-function, noting that $\sum_{i=1}^n\omega_i=-\text{r}$.

Remarkably, expression (\ref{correlator})-(\ref{alpha}) is exactly the one appearing in the generalized $H_3^+$ WZW-Liouville correspondence \cite{R}; specifically, see equations (3.1), (3.5), (3.29), and (3.30) of Ref. \cite{R}, where $\lambda $ here corresponds to $c_k^{-1}$ there. Besides, here we have a sum in $\text{r} $ over $n-2 $ such contributions. In other words, the expression for the correlation function in the theory (\ref{Uno})-(\ref{Deformation}) is exactly the sum of $n-2$ different $n$-point correlation functions in the $SL(2,\mathbb{R})_k$ WZW theory, each of them with a different total winding number $w$; namely
\begin{eqnarray}
&&\left\langle \prod_{i=1}^{n}\,  V_{j_i,m_i,\bar{m}_i}^{\omega_i}(z_i,\bar{z}_i) \, \right\rangle =\, \frac{c_k^2}{2\pi^3b}\, \sum_{w =0}^{n-2} \, \delta\Big(\sum_{i=1}^n\omega_i+w\Big)\, \left\langle \prod_{i=1}^{n}\,  \Phi_{j_i,m_i,\bar{m}_i}^{\omega_i}(z_i,\bar{z}_i) \, \right\rangle^{(w)}_{\text{WZW}}\nonumber
\end{eqnarray}
where $\Phi_{j,m,\bar{m}}^{\omega}(z,\bar{z})$ are vertex operators in the $SL(2,\mathbb{R})_k$ WZW theory that create Kac-Moody primaries of the $\omega$-spectrally flowed $j$-representation of the $\hat{sl}(2)_k$ affine algebra. The label $\omega_i$ corresponds to the winding number of the $i^{\text{th}}$ string state in AdS$_3$. The overall factor ${c_k^2}/({2\pi^3b})$ is independent of $n$ and the momenta, and so it can be absorbed in the definition of the inner product. The superindex $w$ in the correlator on the right hand side indicates that it corresponds to the correlator for the case in which the total winding number is violated in $w $ units. Recall $0\leq w \equiv -\sum_{i=1}^n\omega_i \leq n-2$.

This is compatible with the interpretation proposed in \cite{MO, MO3}, according to which the wave function of a string state in AdS$_3$ can be thought of as a linear combination of operators with well-defined winding number $\omega $, namely $V_{j,m,\bar m}(z,\bar z)=\sum _{\omega}d_{\omega}(k)\,V_{j,m,\bar m}^{\omega }(z,\bar z)$ with $d_{\omega}(k)$ being coefficients, i.e., as the sum of contributions $V_{j,m,\bar m}^{\omega }(z,\bar z)$ creating states of different spectral flow representations of $\hat{sl}(2)_k$.

\section{String amplitudes in AdS$_3$}

\subsection{KPZ scaling and string coupling}

Before moving on to the string theory interpretation of these correlators, let us comment on the Knizhnik-Polyakov-Zamolodchikov (KPZ) scaling of correlators (\ref{correlator}). So far, we have considered the Liouville cosmological constant $\mu$ equal to 1. We have done this without loss of generality as the value of $\mu$ can always be changed by shifting the zero mode of the Liouville field: $\phi_+\to \phi_++c$. Under this, the Liouville cosmological constant gets rescaled as $\mu \to e^{\sqrt 2 b c}\mu $. This transformation also affects the operator $\mathcal{O}(z,\bar z )$, which gets rescaled by a factor $e^{-\frac{c}{\sqrt 2 b}}$. That is to say, we find the relation
\begin{equation}
\frac{\delta \mu}{\mu} = -2b^2\, \frac{\delta \lambda }{\lambda }.
\end{equation}
Then, we find the following parametric relation between constants
\begin{equation}
{\lambda } \, \sim \, \mu^{-\frac{1}{2b^2}} .\label{escaleos}
\end{equation}
This is important because it would allow us to associate the KPZ scaling of the correlators with the dependence on the string coupling constant exhibited by the string scattering amplitudes. First, note that the Liouville correlation functions on the sphere scale like 
\begin{equation}
\left\langle \prod_{l=1}^{n'}\,  V^{L}_{\alpha_+^l}(z_l,\bar{z}_l) \, \right\rangle_{\text{LFT}}\, \sim \, \mu^{\kappa}
\end{equation}
with 
\begin{equation}
 \kappa=\frac{Q_+}{b}-\frac{1}{b}\sum_{l=1}^{n'}\alpha_+^l = 1+\frac{1}{b^2}-\frac 1b \sum_{i=1}^n\alpha_+^i+\frac{1}{2b^2}(n-2-\text{r}) 
\end{equation}
In the case we are dealing with, we have $n'=2n-2-\text{r}$ Liouville vertices, $n-2-\text{r}$ of them having momentum $\alpha_+^{l}=-\frac{1}{2b}$ for $n+1\leq l\leq 2n-2-\text{r}$. On the other hand, we have an overall factor $\lambda ^{-2-\text{r}}$. Taking into account (\ref{escaleos}), this yields the total KPZ scaling 
\begin{equation}
\left\langle \prod_{i=1}^{n}\,  V_{j_i,m_i,\bar{m}_i}^{\omega_i}(z_i,\bar{z}_i) \, \right\rangle\, \sim\, \mu^{\hat \kappa}
\end{equation}
with
\begin{equation}
\hat \kappa = \kappa +\frac{\text{r}+2}{2b^2} = 1+\frac{1}{b^2}-\frac 1b\sum_{i=1}^n\alpha^i_++\frac{n}{2b^2}=1-\sum_{i=1}^{n}(j_i+1),
\end{equation}
which is exactly the way the tree-level string amplitudes in AdS$_3$ scale. This is of course compatible with the fact that the value of $\mu$ can be changed by shifting the zero-mode of $\phi_+$, which, through the background charge, results in nothing but a shifting of the linear dilaton term. In fact, we find the following relation between the Liouville cosmological constant and the (3-dimensional) string coupling constant
\begin{equation}
\mu \, \sim\, g_s^{-2}\, ,\label{gs}
\end{equation}
cf. \cite{Notes}. This is also compatible with the freedom of shifting the zero mode $\phi_-(z,\bar z)\to \phi_-(z,\bar z)+\sqrt{\frac k2}\log \lambda$ to set $\lambda$ to 1. For higher genus the matching also works: at genus-$\text{g}$, one finds $\kappa = (1-\text{g})(1+{b^{-2}})-b^{-1}\sum_{l=1}^{n'}\alpha_+^l$, which yields $\hat{\kappa } = 1-\text{g} - n-\sum_{i=1}^{n}j_i$, and the sum in $\text{r}$ goes up to $n-2+2\text{g}$. Below, we will comment on the tree-level and the one-loop amplitudes.

\subsection{Tree-level amplitudes}

Now, let us discuss how to interpret the results of section 3 from the perspective of string theory on AdS$_3$. Tree-level string scattering amplitudes in AdS$_3\times \mathcal{M}$ are given by integrating the worldsheet correlation functions on the Riemann sphere; namely   
\begin{equation}
\mathcal{A}^{\text{ string}}_{\text{p}_1,\text{p}_2,...,\text{p}_n} \, = \, \int \prod_{i=4}^{n}d^2z_i\,\,
\left\langle \prod_{i=1}^{n}\,  \Phi_{j_i,m_i,\bar{m}_i}^{\omega_i}(z_i,\bar{z}_i) \, \right\rangle_{\text{WZW}}\times \, \, \left\langle \prod_{i=1}^{n}\,  \text{V}^{(\mathcal{M})}_{h_i,\bar{h}_i}(z_i,\bar{z}_i) \, \right\rangle_{\text{CFT}(\mathcal{M})}
\end{equation}
where we have fixed $z_1=0$, $z_2=1$ and $z_3=\infty $ to cancel the volume of the conformal Killing group,
\begin{equation}
\text{Vol}(PSL(2,\mathbb{C}))\, = \,\int \prod _{i=1}^3d^2z_i\, \prod_{i<j}^3|z_{ij}|^{-2}\, ,
\end{equation}
and where we have introduced the momenta-labels $\text{p}_i$ to collectively refer to the indices $j_i,m_i,\bar{m}_i, \omega_i $, as well as to the momenta $h_i,\bar{h}_i$ of the internal CFT$_2$ on $\mathcal{M}$.

Then, considering that the total winding number in AdS$_3$ can be violated, the correlators can be thought of as a sum of contributions with different total winding number; namely
\begin{equation}
\left\langle \prod_{i=1}^{n}\,  \Phi_{j_i,m_i,\bar{m}_i}^{\omega_i}(z_i,\bar{z}_i) \, \right\rangle_{\text{WZW}} = \, \sum_{w =0}^{n-2} \, \delta\Big(\sum_{i=1}^n\omega_i+w\Big)\, \left\langle \prod_{i=1}^{n}\,  \Phi_{j_i,m_i,\bar{m}_i}^{\omega_i}(z_i,\bar{z}_i) \, \right\rangle^{(w)}_{\text{WZW}}
\end{equation}
and, according to what we have shown above, these correlators can be represented by the dual model (\ref{dUno})-(\ref{dFinal}). In other words, we have
\begin{equation}
\mathcal{A}^{\text{ string}}_{\text{p}_1,\text{p}_2,...,\text{p}_n} \, = \, \int \prod_{i=4}^{n}d^2z_i\,\,
\left\langle \prod_{i=1}^{n}\,  V_{j_i,m_i,\bar{m}_i}^{\omega_i}(z_i,\bar{z}_i) \, \right\rangle_{\text{CFT}(\phi_{\pm }, X)}  \times \, \,...
%\left\langle \prod_{i=1}^{n}\,  \text{V}^{(\mathcal{M})}_{h_i,\bar{h}_i}(z_i,\bar{z}_i) \, \right\rangle_{\text{CFT}(\mathcal{M})}\, .\nonumber
\end{equation}
where the ellipsis stands for the internal contribution.

In conclusion, we have proven what we stated at the beginning: the worldsheet CFT$_2$ of string theory on AdS$_3$ can alternatively be represented by a spacelike Liouville theory ($\phi_+$) coupled to a timelike Liouville-like field ($\phi_-$) with (real) background charge, and a free boson ($X$), and deformed by the marginal operator $\mathcal{O}(z,\bar{z})$ in (\ref{Deformation}). The central charge of the spacelike Liouville theory tends to that of the CFT dual, $c_+\simeq 6k$, in the semiclassical limit $k=R^2_{\text{AdS}}/\alpha' \to \infty $.

\subsection{Beyond tree-level}

Before concluding this section, let us make a comment on the one-loop amplitudes in order to illustrate how all this works beyond tree-level. This amounts to consider the CFT$_2$ defined on a torus with modular parameter $\tau$. In that case, and focusing on the case of spinless states $m_i=\bar{m}_i$ for simplicity, one finds an expression similar to (\ref{correlator}), which turns out to be the sum of $n$ terms (from $\text{r}=0$ to $\text{r}=n$) of the following form 
\begin{eqnarray}
&&\frac{1}{\lambda^{\text{r}+2}}\prod_{i=1}^{n}\frac{\Gamma(-j_i-m_i)}{\Gamma(1+j_i+m_i)}\, \prod_{i<i'}^{n}\chi(z_{ii'}|\tau )^{2\beta_{ij}}\, \int \prod_{a=1}^{n}d^2y_{a}
\,
\prod_{a=1}^{n-\text{r}}\prod_{i=1}^{n}\chi(z_i-y_a|\tau )^{2m_i-k}
\, \nonumber \\
&&\ \ \ \ \ \ \ \ \ \ \ \times \,\prod_{a<a'}^{n-\text{r}}\chi(y_{aa'}|\tau )^{k}\,\, \left\langle \prod_{i=1}^n V^L_{\alpha^i_+}(z_i) \, \prod_{a=1}^{n-\text{r}} V^{L}_{-\frac{1}{2b}}(y_a) \,\right\rangle_{\text{LFT}}
\end{eqnarray}
where 
\begin{equation}
\chi(z|\tau )=e^{-\frac{\pi (\text{Im}(z))^2}{\text{Im}(\tau )}}\, \Big| \frac{\vartheta_1(z|\tau )}{\vartheta'_1(0|\tau )}\Big|
\end{equation}
with $\vartheta_1$ being the Jacobi function
\begin{equation}
\vartheta_1(z|\tau )=-\sum_{n=-\infty}^{\infty}e^{i\pi(n+\frac 12)^2\tau +2\pi i(n+\frac 12 )(z+\frac 12 )}
\end{equation}
and $\vartheta'_1$ its derivative with respect to $z$. 

In the case of the 1-loop amplitudes, the KPZ scaling is
\begin{equation}
\sim\, \mu^{-\sum_{i=1}^n(j_i+1)}\, ,
\end{equation}
which is consistent with the identification (\ref{gs}) as well as with the scaling of the vertex operators.

\section{Marginal deformation}\label{Section5}

\subsection{Two marginal operators}

Let us analyze the properties of the deformation operator $\mathcal{O}(z,\bar z )$ in (\ref{Deformation}). In order to do so, it is worth comparing (\ref{dFinal}), namely
\begin{equation}
\mathcal{O}(z,\bar{z})\, \propto\, \mu^{1-\frac{k}{2}} \,\, e^{-\sqrt{\frac{k-2}{2}}\,\phi_+-\sqrt{\frac{k}{2}}\,\phi_-} ,\label{O}
\end{equation}
with the Liouville-wall operator
\begin{equation}
\mathcal{O}_{L}(z,\bar{z})\, = \,\mu \, \, e^{\sqrt{\frac{2}{k-2}}\,\phi_+}\label{Ol}.
\end{equation}

We observe that, unlike $\mathcal{O}_L(z,\bar z )$, operator $\mathcal{O}(z,\bar z )$ depends both on the spacelike Liouville field $\phi_+$ and the timelike field $\phi_-$, defining the direction $\sqrt{\frac{k-2}{2}}\phi_++\sqrt{\frac{k}{2}}\phi_-$ in the target space, which, while always timelike, tends to the lightlike direction $\sim -(\phi_++\phi_-)$ in the semiclassical limit $k\to \infty$. In this sense, this behaves like a tachyonic shock wave. 

Operator $\mathcal{O}(z,\bar z )$ grows exponentially in the direction $\phi_+\to -\infty$. This is in sharp contrast to the Liouville operator $\mathcal{O}_L(z,\bar z )$, which grows exponentially in the opposite direction, $\phi_+\to +\infty$. This means that one becomes important deep into the bulk, while the other does so near the boundary. In order to see this, it is worth resorting to the Wakimoto representation of the $\hat{sl}(2)_k$ symmetry algebra, which consists of a dimension-($1,0$) ($\beta, \gamma$) ghost system plus a scalar $\varphi$ with background charge $Q_{\varphi}=-1/\sqrt{k-2}$. This well-known realization of the worldsheet theory in AdS$_3$ exhibits a self-interaction potential $\beta \bar \beta e^{-\sqrt{\frac{2}{k-2}}\varphi}$ with the boundary of AdS$_3$ being at $\varphi=\infty$. The Liouville direction in (\ref{Ol}) would effectively correspond to $\phi_+=-\varphi$, cf. \cite{GN3}.

Another crucial difference between operators $\mathcal{O}(z,\bar z )$ and $\mathcal{O}_L(z,\bar z )$ is their behavior at large $k$. While in the semiclassical limit, $k\to \infty$, the Liouville operator (\ref{Ol}) becomes a steep ramp that acts as a rigid wall, operator (\ref{O}) becomes suppressed. This behavior is reflected in the relation between the couplings
\begin{equation}
\lambda^{-\sqrt{\frac{2}{k-2}}} \, \sim\, \mu^{\sqrt{\frac{k-2}{2}}}\, .
\end{equation}
This might resemble the Langlands duality between $k-2\leftrightarrow \frac{1}{k-2}$ in $SL(2,\mathbb{R})_k$ WZW, which somehow relates to the Liouville self-duality under $b\leftrightarrow\frac 1b$; however, it is not exactly that: the Liouville operator that constitutes (\ref{O}) is the degenerate operator $V_{-\frac{1}{2b}}^{L}$, which, unlike the dual Liouville screening operator $V_{+\frac {1}{b}}^L$, has a negative sign in the exponent, so being relevant in the opposite Liouville direction.

\subsection{Marginal operator in the sector $\omega =-2$}

The role of operator $\mathcal{O}(z,\bar z )$ is to change the winding number of the string states. We have seen this explicitly when deriving the correlation functions: the introduction of $\ell $ such operators in the correlator leads to the violation of the winding number $w$ in $n-2-\ell $ units (with $0\leq \ell \leq n-2$). Such a process occurs deep into the bulk. This interpretation is reinforced by looking at the quantum numbers of the state that operator $\mathcal{O}(z,\bar z )$ would correspond to. In fact, operator (\ref{Ol}) represents a state of momenta
\begin{equation}
\alpha_+=-\frac{1}{2b}, \ \ \ \ \alpha_-=\bar{\alpha}_-=-{\frac{\sqrt{k}}{2}} , \ \ \ \ p=\bar p = 0 \, ,
\end{equation}
which, in terms of the WZW natural variables, read
\begin{equation}
j=1-k, \ \ \ \ m=\bar{m}=k, \ \ \ \ \omega =-2. \label{L54}
\end{equation}
Firstly, we observe that this yields the conformal dimensions
\begin{equation}
\Delta_{1-k,k,-2}=\bar \Delta_{1-k,k,-2}=1\,,\label{marginality}
\end{equation}
which means that the operator $\mathcal{O}(z,\bar z )$ is marginal. In fact, it can be seen that $\mathcal{O}(z,\bar z )$ is {\it exactly} marginal. In addition, it yields zero energy,
\begin{equation}
m+\frac k2 \omega = \bar m+\frac k2 \omega = 0
\end{equation}
Secondly, we observe that $\mathcal{O}(z,\bar z )$ belongs to the spectral flow sector $|\omega |=2$. This represents an important difference from the previous realization proposed in \cite{G}, where a description of the AdS$_3$ string worldsheet based on (sine-)Liouville theory was proposed.

\subsection{Marginal operator in the sector $\omega =-1$}

In the realization proposed in \cite{G} the perturbation is given by an operator of the sector $|\omega |=1$. More precisely, the deformation operator considered in \cite{G} corresponds to momenta 
\begin{equation}
j=1-\frac k2 , \ \ \ \ m=\bar m = \frac  k2, \ \ \ \  \omega =-1\, ,
\end{equation}
which also yields 
\begin{equation}
\Delta_{1-\frac k2, \frac k2 , -1}=\bar \Delta_{1-\frac k2 , \frac k2, -1}=1
\end{equation}
with
\begin{equation}
m+\frac k2 \omega = \bar m+\frac k2 \omega = 0\, .
\end{equation}
Note that, in the conventions here, this $\omega =-1$ operator is nothing but $\mathcal{O}_L(z,\bar z )$, with values of momenta
\begin{equation}
\alpha_+=b, \ \ \ \ \alpha_-=\bar{\alpha}_-=0 , \ \ \ \ p=\bar p = 0 \, ,
\end{equation}
which corresponds to $j=1-\frac k2<0$, namely $\Phi_{1-\frac k2 , \frac k2 , \frac k2}^{-1}=e^{\sqrt{2}b\phi_+}$.

\subsection{The spectral flow operator}

Through the so-called spectral flow duality, which relates discrete representations $\mathcal{D}^{\pm ,\omega \mp 1}_{j}$ with highest- (resp. lowest-) states of representations $\mathcal{D}^{\mp ,\omega }_{-\frac k2 -j}$, and considering the Weyl reflection symmetry under $j\leftrightarrow -1-j$, operator $\mathcal{O}(z,\bar z )$ can be associated to a state with momenta $j=-\frac k2$, $m=\bar m = \pm \frac k2$, $\omega = \mp 1$. The latter is exactly the spectral flow operator $\Phi^{\pm 1}_{-\frac k2, \mp \frac k2, \mp \frac k2}$ introduced in \cite{MO3}, whose insertion in the operators produces the violation of the winding number. Note that the map
\begin{equation}
j\, \to \, \tilde j =-\frac k2-j\, , \ \  \ \ \ m\, \to \, \tilde m = \frac k2+m\, , \ \ \ \ \  \omega \, \to \, \tilde \omega = \omega -1 
\end{equation}
transforms $\Delta_{j,m,\omega}$ as follows
\begin{equation}
\Delta_{j,m,\omega} \, \to \, \tilde \Delta_{\tilde j,\tilde m,\tilde \omega}=\Delta_{j,m,\omega}-j-m
\end{equation}
while keeping
\begin{equation}
\tilde m+\frac k2 \tilde\omega \, = \, m+\frac k2 \omega \, .
\end{equation}
Therefore, for highest-weight states with $j+m=0$ this corresponds to a symmetry of the spectrum, namely 
\begin{equation}
\tilde \Delta_{-\frac k2 -j,-\frac k2 - j, \omega }=\Delta_{j,\, j,\, \omega +1}\, .
\end{equation}
In contrast, for $j=1-k$, $m=k$, we have 
\begin{equation}
1+\tilde \Delta_{-\frac k2,\frac k2 , -1}=\Delta_{1-k,k,-2}\, .
\end{equation}
Indeed, unlike both operators $\Phi^{-1}_{1-\frac k2, \frac k2, \frac k2}$ and $\Phi^{- 2}_{1- k,  k, k}$, which have conformal dimension 1, the spectral flow operator $\Phi^{\pm 1}_{-\frac k2, \mp \frac k2, \mp \frac k2}$ has a vanishing conformal dimension 
\begin{equation}
\Delta_{-\frac k2, \mp \frac k2, \pm 1}=\bar \Delta_{-\frac k2, \mp \frac k2, \pm 1}=0
\end{equation}
with
\begin{equation}
m+\frac k2 \omega =\bar m+\frac k2 \omega =0
\, ;\end{equation}
this is why there are also those who refer to the latter as the conjugate representation of the identity.  

\subsection{Remarks}

%Some remarks are in order:

It may not have gone unnoticed that the operator $\mathcal{O}(z,\bar z )\propto e^{-\sqrt{\frac{k-2}{2}}\phi} e^{- i\sqrt{\frac{k}{2}} \theta}$ resembles the two marginal operators that constitute the action of sine-Liouville theory, a theory which, as has been known since the work of Fateev, Zamolodchikov and Zamolodchikov (FZZ), is dual to the $SL(2,\mathbb{R})/U(1)$ WZW model; see \cite{KKK}. The sine-Liouville operator contains the operators $\mathcal{O}_{\pm } (z,\bar z )\propto e^{-\sqrt{\frac{k-2}{2}}\phi} e^{+ i\sqrt{\frac{k}{2}}\tilde \theta}+e^{-\sqrt{\frac{k-2}{2}}\phi} e^{- i\sqrt{\frac{k}{2}}\tilde \theta}$, where $\tilde \theta(z,\bar z)= i\phi_-(z)-i\bar \phi_-(\bar z)$. In fact, the values $\alpha_+=-\frac{\sqrt{k-2}}{2}$, $\alpha _-=  \frac{\sqrt{k}}{2}$, and $p=\bar p = 0$ are exactly those assumed by the fields in sine-Liouville theory. Nevertheless, the two models are not exactly identical; in particular, here we have a single operator $\mathcal{O}(z,\bar z )$, and it depends on $\theta (z,\bar z)= i\phi_-(z)+i\bar \phi_-(\bar z)$ rather than on its T-dual combination of modes. Besides, in sine-Liouville theory only the Liouville-type field $\phi(z,\bar z)=\phi_+(z)+\bar \phi_+(\bar z)$ has a non-vanishing background charge. The relationship between this type of representation was studied in \cite{Gaston2}; cf. \cite{G}.

The second remark is about self-duality. There is a second operator with $\omega =-1$ and $m=\frac k2$; it corresponds to the dual screening operator in Liouville theory, $\Phi_{\frac k2 - 2, \frac k2 , \frac k2}^{-1}=V_L=e^{\frac{\sqrt{2}}{b}\phi_+}$, with $j=\frac k2-2>-1$. In other words, the Liouville self-duality $b\leftrightarrow \frac 1b$ in this case corresponds to the Weyl reflection $j\leftrightarrow -1-j$.

Lastly, let us go back to the Liouville operator $\mathcal{O}(z,\bar z )$, whose momentum is $\alpha_{+}=-\frac{1}{2b}$. This value of the momentum is special: it marks the first non-trivial degenerate state in Liouville theory; that is, the state containing a null state within the Virasoro Verma module at the lowest level. It is, therefore, natural to ask whether there might be other operators of interest corresponding to other degenerate states. These values correspond to $\alpha_+ = -\frac{r}{2}b-\frac{s}{2}b^{-1}$, with $r,s\in \mathbb{Z}_{\geq 0}$. Although there are no particularly interesting marginal operators associated with these values of $\alpha_+$ for $r>2$, it is worth noting that, in terms of $j$, these values of $\alpha_+$ take the form: $j = -\frac{r+2}{2}-\frac{s+1}{2}(k-2)$. These latter values correspond to the $SL(2,\mathbb{R})_k$ momenta of the so-called Kac-Kazhdan series, which admit highest-weight representations within the $\hat{sl}(2)_k$ affine Kac-Moody algebra. This is significant as it suggests a means of establishing a direct link between the Virasoro algebra and the affine algebra.

\subsection{Comments on the orbifold deformation}

The worldsheet realization based on Liouville theory that we propose here has, of course, a very different meaning from the orbifold CFT$_2$ construction, even though the latter also involves Liouville theory. Nevertheless, there are certain similarities between the two theories that are worth highlighting.

Our CFT construction involves a relation between the Kac-Moody level $k$ and the Liouville theory parameter $b$ that is the same as the one appearing in the lineal dilaton (Liouville-type) description of the dual theory; namely, $b = \frac{1}{\sqrt{k-2}}$. However, this should not mislead us: the relation between $b$ and the background charge in both realizations is actually very different: in our case, the Liouville background charge is $Q_+=b+\frac 1b =\frac{k-1}{\sqrt{k-2}}$ rather than the value $Q=-b+\frac 1b =\frac{k-3}{\sqrt{k-2}}$ considered in \cite{Eberhardt2}. The background charge $Q_+$ (the one with the plus sign) is the well-known result in the $c\geq 25$ Liouville theory, and appears, for example, in the $H_3^+$ WZW-Liouville correspondence (see Eq. (2.2) in Ref. \cite{RT} and the discussion around (2.19)-(2.20) therein; see also the discussion around Eqs. (3.4)-(3.5) in Ref. \cite{R}, and the relation between $Q_{\phi}$ and $Q_{\varphi}$ shown in pages 4 and 6 of Ref. \cite{Schomerus}). On the other hand, the background charge $Q$ (the one with the minus sign) appears in the effective description of the theory on a single long string \cite{SeibergWitten} --in Appendix A we briefly review how this charge appears in the CFT$_2$ computation. 

The discrepancy between $Q_+=\frac{k-1}{\sqrt{k-2}}$ and $Q=\frac{k-3}{\sqrt{k-2}}$ should not surprise us, since, as we said, the two Liouville descriptions have very different meanings. Notice, however, that both Liouville constructions share some similarities. In particular, in both cases the large $k$ limit of the Liouville central charge is the same, namely $c\simeq 6k+\mathcal{O}(1)$. This is because, in the semiclassical limit, both $Q_+$ and $Q$ grow parametrically like $\sim\sqrt{ k}$, and the main contribution to the central charge actually comes from the background charge. The finite-$k$ regime, on the other hand, is quite different: while the background charge $Q_+\in \mathbb{R}_{\geq 2}$ yields a central charge $c_+ \geq 25$, the background charge $Q\in \mathbb{R}$ yields a central charge $c\geq 1$. This leads to a significant difference when analyzing the tensionless point, which in the bosonic string is $k=3$. In that case, the Liouville contribution to the worldsheet theory yields $c_+=25$, whereas the linear dilaton theory appearing in the orbifold construction yields the expected value $c=1$.

The difference between the background charges in the two Liouville realizations leads to different relations between the Liouville momenta $\alpha$ and the isospin variables $j$. This is further compounded by the different sign conventions adopted in the various papers (for example, $j$ in \cite{Eberhardt2} is our $-j$). After standardizing the notation, the different relations between $\alpha$ and $j$ that appear in the literature are as follows: while for us $\alpha=b(\frac k2 +j)$, in \cite{Eberhardt2} the authors have $\alpha=b(\frac k2-j-2)$ (see Eq. (2.6) there); in \cite{231205317}, on the other hand, the relation is $\alpha=b(-\frac k2 -j+1)$ (see Eq. (4.4) there); see also \cite{Knighton:2024qxd}. Due to these differences, when comparing the operators that appear in the various realizations, it is convenient to refer to the values of $\alpha $ instead of $j$. When doing so, one notes that the Liouville momentum associated to the twist operator $\sigma_{2,-\frac{1}{2b}}(x,\bar x)$ introduced in \cite{Eberhardt2} is $\alpha=-\frac{1}{2b}$ (see Eq. (2.46) therein), which is exactly the same value $\alpha_+=-\frac{1}{2b}$ of our marginal operator $\mathcal{O}(z,\bar z)$. That is, both in the realization of the orbifold CFT$_2$ \cite{Eberhardt2} and in our worldsheet construction, the marginal deformation includes a Liouville dressing whose momentum is that of a degenerate field. This leads us to wonder whether there might be a more direct connection between the twist operators $\sigma_{2,-\frac{1}{2b}}(x,\bar x)$ of \cite{Eberhardt2} and our operator $\mathcal{O}(z,\bar z)$. We cannot answer this question here, given that the precise relationship between the two constructions is something we do not yet fully understand (see Appendix B). We hope to address this question in the near future. For the moment, we can be content to point out some extra similarities between our Liouville realization of the worldsheet and the orbifold CFT$_2$ construction. For example, we note that our marginal operator $\mathcal{O}_L(z,\bar z)$ is also similar to an operator that appears in the analysis of the orbifold CFT$_2$ in Ref. \cite{231205317}. After matching the sign conventions, it is possible to observe that the operator $\Phi^{\frac{1}{2b^2},-1}_{\frac k2 , \frac k2}(v,\bar v)$ appearing in Eq. (4.8) of Ref. \cite{231205317} corresponds to our marginal operator $\mathcal{O}_L(z,\bar z)$, with $j=1-\frac k2$, $\omega =-1$, and $m=\bar m=\frac k2$; cf. Eq. (4.21) of \cite{231205317}. It may also be interesting to compare our operators $\mathcal{O}(z,\bar z)\propto V_{1-k,k,k}^{-2}$ and $\mathcal{O}_L(z,\bar z)\propto V_{1-\frac k2,\frac k2,\frac k2}^{-1}$ with the operators $\Phi^{(-2)}_{k;k,k}$ and $\Phi^{(-1)}_{1-\frac k2;\frac k2-1,\frac k2 -1}$ that appear in the construction of Refs. \cite{Kutasov} and \cite{Balthazar:2021xeh} (such a comparison requires taking into account, in addition to the different sign conventions, the shift $k\to k-2$ when moving from the supersymmetric theory to the bosonic theory). We also plan to analyze the relationship with \cite{Balthazar:2021xeh} in the near future.

\section{Conclusions}

We propose an alternative worldsheet description of bosonic string theory on AdS$_3$ supported by NS--NS flux. The theory is formulated in terms of a spacelike Liouville field with central charge $c\geq 25$, a timelike scalar field carrying a real background charge, and a free boson, together with a marginal deformation. The resulting conformal field theory has the same total central charge as the SL$(2,\mathbb{R})_k$ Wess-Zumino-Witten model and reproduces the spectrum of string states in AdS$_3$, including the sectors associated with spectral flow and winding strings. In the semiclassical limit, the Liouville sector acquires a central charge $c\simeq 6k$, matching the central charge of the holographic dual conformal field theory.

We compute the correlation functions of the proposed worldsheet theory and show that they admit an exact representation in terms of Liouville correlation functions with insertions of degenerate operators. After integrating out the auxiliary fields, the resulting expressions coincide with those appearing in the generalized H$_3^+$ WZW--Liouville correspondence. This establishes that the correlation functions of the deformed Liouville theory reproduce the scattering amplitudes of strings propagating in AdS$_3$. In particular, the expansion in powers of the marginal deformation naturally generates contributions with different amounts of winding number violation, yielding precisely the pattern known from the SL$(2,\mathbb{R})_k$ description. The dependence of the amplitudes on the Liouville cosmological constant is also shown to reproduce the expected KPZ scaling and its interpretation in terms of the string coupling constant.

A central role in this construction is played by the marginal deformation operator. We show that it corresponds to a non-normalizable state belonging to the spectral-flow sector $\omega=-2$, carrying vanishing spacetime energy and generating winding number violation in scattering processes. This operator differs substantially from the Liouville interaction usually associated with the Wakimoto representation and from the $\omega=-1$ deformation that appears in earlier sine-Liouville-inspired descriptions of the AdS$_3$ worldsheet theory. Its properties reveal a new realization of the string dynamics deep in the AdS$_3$ bulk and provide a direct worldsheet mechanism for winding number changing processes.

Finally, we compare our construction with recent proposals in which Liouville theory appears in the holographic dual description of AdS$_3$ string theory, particularly in deformed symmetric-orbifold models. Although the two Liouville realizations have different physical interpretations and involve different background charges, both contain marginal deformations dressed by the same degenerate Liouville momentum $\alpha=-1/(2b)$. This striking similarity motivates a closer examination of the relation between worldsheet and spacetime Liouville descriptions. While no direct connection is established here, the parallel structures uncovered by our analysis suggest that the Liouville formulation of the AdS$_3$ worldsheet may provide new insight.

The possibility of a closer relationship between these two Liouville constructions has already been suggested in the literature. In Ref. \cite{Dei:2022pkr}, the authors observed that the holographic duality analyzed in their paper relates string correlation functions of the $SL(2,\mathbb{R})_k$ WZW model to correlation functions of a deformed symmetric orbifold, and that the $H_3^+$ WZW-Liouville correspondence is similar in spirit. They noted that it is somewhat puzzling that the latter relation does not play a role in their analysis and that it would be very interesting to understand the connection between these two statements. The $H_3^+$ WZW-Liouville correspondence does play an important role in our derivation of the worldsheet correlators, and that is the reason why we hope our analysis can help to better understand the relationship between the two realizations. We hope to return to this issue in the future.

\subsection*{Acknowledgments}

G.G. thanks Lorenz Eberhardt and Massimo Porrati for discussions.

\appendix

\section{On a long string}

Let us briefly review the construction of the effective theory on a single long string, as first introduced in \cite{SeibergWitten}. This construction forms the basis of the dual description of string theory on AdS$_3 \times \mathcal{M}$ in terms of the orbifold CFT$_2$.

The physical degrees of freedom living on the long string consist of the modes of the $\sigma $-model on $\mathcal{M}$, describing the location of the string in the integral manifold, together with a field that parameterizes the radial direction in AdS$_3$. The latter is given by a Liouville-type field.

The construction of the long string theory involves a twist in the stress tensor. Evaluated on the classical string solution that describes a single winding long string near the boundary, the $SL(2,\mathbb{R})_k$ currents yield $J^-=J^3=0$ and $J^+=\text{const}$. Let us review how it works explicitly: a solution describing $\omega $ coincident long strings expressed in Wakimoto fields is $\gamma(z)=q(z-z_i)^{\omega}$ with $q\in \mathbb{R}$, $z_i\in \mathbb{C}$, and $\omega\in\mathbb{Z}_{\geq 0}$ being three constants; $\varphi=\varphi_0$ is also a constant. Analogously, we have $\bar\gamma(\bar z)=q(\bar z-\bar z_i)^{\omega}$. This corresponds to a holomorphic solution ($\bar \partial \gamma=\partial\bar \gamma =0$) that, provided $\varphi_0\gg \sqrt{k}$, locates near the boundary. Near the boundary, the effective potential $\beta\bar \beta e^{-\sqrt{\frac{2}{k-2}}\varphi}$ vanishes. On the other hand, integrating the auxiliary fields $\beta $ and $\bar \beta $ the WZW action acquires an interaction term $\bar\partial \gamma \partial \bar \gamma e^{\sqrt{\frac{2}{k-2}}\varphi}$ that turns out to be transparent for a holomorphic configuration, so it does not prevent the long strings from extending all the way to the boundary. The winding number is computed by $\oint_{S_i} dz\,\partial\gamma/\gamma=2\pi i\, \omega$, where $S_i$ is a contour that encircles the point $z_i$. The single long string case is $\omega =1$ and, in fact, yields a constant value $J^+=qk$. Then, by analyzing how the spacetime stress tensor acts on the BRST cohomology and demanding $J^+$ to have conformal dimension zero, one is led to improve the stress tensor as follows
\begin{equation}\label{improva}
T_{SL(2,\mathbb{R})}(z)\, \to \,  \hat T(z)\, =\, T_{SL(2,\mathbb{R})}(z) +\partial J^3(z)\, ,
\end{equation}
This produces the improved central charge
\begin{equation}
c_{SL(2,\mathbb{R})}\,\to \, \hat{c}\,=\, c_{SL(2,\mathbb{R})}+6k= 3+6\Big(\,\frac{1}{\sqrt{k-2}}+\sqrt{k-2}\,\Big)^2    \label{deltac}
\end{equation}
which follows from the OPE
\begin{equation}
\partial J^3(z)\,\partial J^3(0) \, = \,\frac k2 \,  \partial^2X(z)\, \partial^2X(0)\, =\, \frac{3k}{z^4} \, + \, ...   
\end{equation}
where we have used the propagator $\langle X(x)\, X(0)\rangle =- \log z$. 

The spacetime stress tensor of the theory on the long string is obtained by adding to $\hat{T}$ the other two contributions of the worldsheet stress tensor, namely $T_{\mathcal{M}}+T_{bc \text{ ghosts}}$. The total stress tensor of the worldsheet, on the other hand, is given by
\begin{equation}
T_{SL(2,\mathbb{R})}+T_{\mathcal{M}}+T_{bc \text{ ghosts}}
\end{equation}
which yields the criticality condition
\begin{equation}
\frac{3k}{k-2}+c_{\mathcal{M}}-26=0 \, .\label{criticality}
\end{equation}
Therefore, one finds that the spacetime central charge is
\begin{equation}
 c= 6k    \, .
\end{equation}
This can be obtained by considering the theory on $\mathcal{M}$ and a Liouville-like field with background charge $Q$ obeying
\begin{equation}
c_{\mathcal{M}}+1+6Q^2=6k
\end{equation}
Using (\ref{criticality}), one obtains
\begin{equation}
Q^2=\frac{(k-3)^2}{k-2}
\end{equation}
and one can finally write 
\begin{equation}
Q=\frac{1}{b}-b\,\in \,\mathbb{R}   \ \ \ \text{with }\ \  b=\frac{1}{k-2}\, .\label{CARGOTA}
\end{equation}
Again, it is important not to mistake (\ref{CARGOTA}) for the background charge in (\ref{deltac}), namely
\begin{equation}
Q_+=\frac{1}{b}+b\,\in \,\mathbb{R}_{ \geq 2}    \ \ \ \text{with }\ \  b=\frac{1}{k-2}\, ;
\end{equation}
this is exactly the discussion we had in section 5.

The OPE between the improved stress tensor $\hat{T}(z)$ and the vertex operators $V_{j,m,\bar m }(z,\bar z )$ is
\begin{equation}
\hat{T}(z)\, V_{j,m,\bar m }(0)\, = \, \frac{\Delta_{j,m,\bar m}+m+\frac k2 \omega }{z^2}\, V_{j,m,\bar m }(0)\, +\, \frac{1}{z}\, \partial V_{j,m,\bar m }(0)\, +\, ... 
\end{equation}
as from (\ref{J3}) we have
\begin{equation}
\sqrt{\frac k2} \partial^2 X(z)\, V_{j,m,\bar m }(0)\, = \, \frac{m+\frac k2 \omega }{z^2}\, V_{j,m,\bar m }(0)\, +\, \frac{1}{z}\, \partial V_{j,m,\bar m }(0)\, +\, ... 
\end{equation}

\section{Deformed symmetric product}

The dual CFT$_2$ proposed in \cite{Eberhardt2} begins by considering a non-normalizable deformation of the symmetric orbifold on the seed $\mathbb{R}_{{Q}}\times \mathcal{M}$, with $\mathbb{R}_{{Q}}$ being a free field $\phi (z,\bar z)$ with the background charge ${Q}\,=\,b^{-1}-b$, namely
\begin{equation}
    S[\phi]\,=\,\frac{1}{4\pi}\int\,d^2z\,\left(\partial\phi\,\bar{\partial}\phi\,+\,{Q}R\phi\right)\,.\label{tucumanos}
\end{equation}
In terms of $k=2+b^{-2}$, the central charge (\ref{CARGOTA}) reads
\begin{equation}
    c\,=\,1\,+\,6{Q}^2\,=\,1\,+\,\frac{6(k-3)^2}{k-2}\,.
\end{equation}
As explained in Appendix A, the central charge of the spacetime stress tensor comes from adding the contribution from the integral theory on $\mathcal{M}$ and taking into account the criticality condition in the worldsheet CFT$_2$, resulting in 
\begin{equation}
    1\,+\,\frac{6(k-3)^2}{k-2}\,+\,\left(26\,-\,\frac{3k}{k-2}\right)\,=\,6k\,.
\end{equation}

The primary operators of the seed theory have the form $e^{\sqrt{2}\alpha\phi}\times V_{\mathcal{M}}$ and create states of conformal weights
\begin{equation}
    h\,=\,\alpha\left({Q}\,-\,\alpha\right)\,+\,h_{\mathcal{M}}\,,\qquad \bar{h}\,=\,\alpha \left({Q}\,-\,\alpha\right)\,+\,\bar{h}_{\mathcal{M}}\,.
\end{equation}

In the symmetric orbifold, not only one has $N$ copies of such a theory, but also has to gauge the permutations $S_N$ that interchange the copies. The spectrum contains both twisted and untwisted sectors. This naturally leads to the introduction of single cycle twist fields, labeled by the twist length $\omega\in\mathbb{Z}_{\geq 1}$ and the weight $k(\omega^2-1)/(4\omega)$. Therefore for every vertex operator in the seed CFT$_2$ with weights $(h-\bar{h})\in\omega\mathbb{Z}$, there is an operator $\sigma_{\omega,\alpha }$ in the twist-$\omega$ sector with weights
\begin{equation}
    h_{\omega}\,=\,\frac{h}{\omega}\,+\,\frac{k\left(\omega^2-1\right)}{4\omega}\,,\qquad \bar{h}_{\omega}\,=\,\frac{\bar{h}}{\omega}\,+\,\frac{k\left(\omega^2-1\right)}{4\omega}\,.
\end{equation}
In terms of the momentum $\alpha$, ignoring the contribution from $\mathcal{M}$, this reads
\begin{equation}\label{eq:weight-twist-field}
    h_{\omega}\,=\,\bar{h}_{\omega}\,=\,\frac{\alpha \left({Q}-\alpha \right)}{\omega}\,+\,\frac{k\left(\omega^2-1\right)}{4\omega}\,.
\end{equation}
From this and the Virasoro constraint on the conformal weights in the worldsheet theory, the relationship between the Liouville momentum $\alpha$ of the twist operator and the isospin $j$ is deduced, namely,
\begin{equation}
    \alpha\,=\,b\left(\frac k2 -j-2\right)
\end{equation}
see Eq. (2.6) in \cite{Eberhardt2}.

An important element in the construction of \cite{Eberhardt2} is the twist-2 operator ($\omega=2$) with Liouville momentum $\alpha=-\frac{1}{2b}$. This yields $h_{2}=\hat{h}_2=1$. Its integrated version is\begin{equation}
    \frac{\lambda}{4\pi}\,\int d^2z\,\sigma_{2,-\frac{1}{2b}}\,\label{lamarginalia}
\end{equation}
with $\lambda\in \mathbb{R}$.

This operator enables us to think of the partially integrated $n$-point correlation functions of the form
\begin{equation}\label{eq:order-m-residue-orbifold}
    \frac{\left(-\lambda\right)^{\ell}}{{\ell}!}\int\prod_{a=1}^{\ell} d^2\xi_a\,\left<\prod_{i=1}^n \sigma_{\omega_i,\alpha_i}(x_i)\,\prod_{a=1}^{\ell} \sigma_{2,-\frac{1}{2b}}(\xi_{a})\right>
\end{equation}
with 
\begin{equation}
\frac{\ell}{2b}\,=\,\sum_{i=1}^n\alpha_i\,-\,Q\label{conditiona}
\end{equation}
as being the residue of $n$-point correlation functions 
\begin{equation}
    \left<\prod_{i=1}^n \sigma_{\omega_i,\alpha_i}(x_i)\right>_{\lambda}\label{uyi}
\end{equation}
defined in the linear dilaton theory (i.e., Liouville theory with vanishing cosmological constant $\mu=0$) deformed by the marginal operator $\sigma_{2,-\frac{1}{2b}}$. That is to say, correlators (\ref{uyi}) are defined in the CFT$_2$ that results from adding (\ref{lamarginalia}) to the linear dilaton action (\ref{tucumanos}). 

It is worth noticing that condition (\ref{conditiona}), expressed in terms of the $SL(2,\mathbb{R})$ variables $j_i$, reads
\begin{equation}
\frac 12\, ({\ell}+2-n)\,{(k-2)}+n-1+\sum_i^nj_i=0\label{dfg}
\end{equation}
which exactly matches the condition for the Liouville correlators (\ref{la311}) not to vanish limit in the limit $\mu=0$. To be more precise, in comparing (\ref{dfg}) with (\ref{la311}) one has to recall that $\text{r}=n-2-\ell$ and notice that the field $\sigma _{2,-\frac{1}{2b}}(x,\bar x)$ is in the sector $\omega =+2$ while the operator $\mathcal{O}(z,\bar z)$ is defined with the convention $\omega =-2$. This makes the relation between marginal operators $\sigma _{2,-\frac{1}{2b}}(x,\bar x)$ and $\mathcal{O}(z,\bar z)$ more evident. 

In \cite{Eberhardt2} the correlation functions (\ref{uyi}) were computed on the sphere for $n=2$ and $n=3$, and the result was shown to match the string amplitudes in AdS$_3$. Throughout such a calculation, the symmetric product correlators can be expressed as a sum over the connected covering surfaces. The number $N$ of sheets of the covering surface of genus $\text{g}$ with a meromorphic map $\gamma$, is set by the Riemann-Hurwitz formula
\begin{equation}
   2 N\,=\,2\,-n\,-\,2\text{g}\,+\,\sum_{i=1}^n\omega_i \,+\,{\ell}\,.
\end{equation}
This equation, with $N$ being the number of screenings in Liouville correlators, is obtained by performing in our CFT the authomorphism $m, \omega \to -m, -\omega $ in both the vertices and screening operators. This makes operator $\mathcal{O}_L(z,\bar z)$ to acquire a dependence on $\phi_-(z,\bar z)$ that consequently changes the condition coming from the integration over the zero mode into  
\begin{equation}
   2 s\,=\,2\,-n\,-\,2\text{g}\,+\,\sum_{i=1}^n\omega_i \,+\,{\ell}\,.
\end{equation}
The number of maps is thus related to the label $\text{r}$ we defined in section 3; see the paragraph between Eqs. (\ref{correlatoru}) and (\ref{correlatoru}); see also the comment below Eq. (\ref{gs}) in section 4.1. Considering the change $\omega_i\to -\omega_i$ to compare with \cite{Eberhardt2}, the variable $\text{r}$ is
\begin{equation}
\text{r}\,=\,\sum_{i=1}^n\omega_i\,\leq\,n-2+2\text{g} \, .
\end{equation}
This allows us to relate the number of sheets of the covering surface ($N$) to the number of insertions of degenerate operators ($\ell$) that must be added to the Liouville factor of the worldsheet theory to produce a given total winding number ($\text{r}$) in the string amplitude on AdS$_3$.

\end{document}